\documentclass[conference]{IEEEtran}
\IEEEoverridecommandlockouts
\usepackage{cite}
\usepackage{amsmath,amssymb,amsfonts}
\usepackage{algorithmic}
\usepackage{graphicx}
\usepackage{textcomp}
\usepackage{xcolor}
\usepackage{todonotes}
\usepackage{multirow}

\usepackage{subfigure}

\usepackage{xcolor}
\def\BibTeX{{\rm B\kern-.05em{\sc i\kern-.025em b}\kern-.08em
    T\kern-.1667em\lower.7ex\hbox{E}\kern-.125emX}}
    
\begin{document}

\newcommand\copyrighttext{%
  \footnotesize This work is sent for a review at the IEEE conference, 2023}
\newcommand\copyrightnotice{%
\begin{tikzpicture}[remember picture,overlay]
\node[anchor=south,yshift=10pt] at (current page.south) {\copyrighttext};
\end{tikzpicture}%
}

\newpage

\makeatletter 
\newcommand{\linebreakand}{%
  \end{@IEEEauthorhalign}
  \hfill\mbox{}\par
  \mbox{}\hfill\begin{@IEEEauthorhalign}
}
\makeatother 

\title{Compressor-Based Classification for Atrial Fibrillation Detection}


\author{\\
\IEEEauthorblockN{NIKITA MARKOV}
\and
\\
\IEEEauthorblockN{KONSTANTIN USHENIN}
\linebreakand
\\
\IEEEauthorblockN{YAKOV BOZHKO}

\and
\\
\IEEEauthorblockN{OLGA SOLOVYOVA}
\\
\\
}

\maketitle
\copyrightnotice

\begin{abstract}
Atrial fibrillation (AF) is one of the most common arrhythmias with challenging public health implications. Therefore, automatic detection of AF episodes on ECG is one of the essential tasks in biomedical engineering. 

In this paper, we applied the recently introduced method of compressor-based text classification with gzip algorithm for AF detection (binary classification between heart rhythms). We investigated the normalized compression distance applied to RR-interval and $\Delta$RR-interval sequences ($\Delta$RR-interval is the difference between subsequent RR-intervals). Here, the configuration of the k-nearest neighbour classifier, an optimal window length, and the choice of data types for compression were analyzed. We achieved good classification results while learning on the full MIT-BIH Atrial Fibrillation database, close to the best specialized AF detection algorithms (avg. sensitivity = 97.1\%, avg. specificity = 91.7\%, best sensitivity of 99.8\%, best specificity of 97.6\% with fivefold cross-validation). In addition, we evaluated the classification performance under the few-shot learning setting. Our results suggest that gzip compression-based classification, originally proposed for texts, is suitable for biomedical data and quantized continuous stochastic sequences in general.
\end{abstract}

\section{Introduction}
Atrial fibrillation (AF) is one of the most widespread arrhythmias with challenging epidemiological consequences. The prevalence of AF continues to increase every year with aging population \cite{staerk2017atrial}, creating a strain on the public healthcare systems. Conversely, AF is characterized by asymptomatic and episodic nature of the disease, leaving the patients unaware of their gradual electrophysiological deterioration. Early detection of arrhythmia paroxysms is the key to patients' successful treatment and improved quality of life.

\begin{figure}[t]
    \centering
    \subfigure
    {
        \includegraphics[width=3.4in]{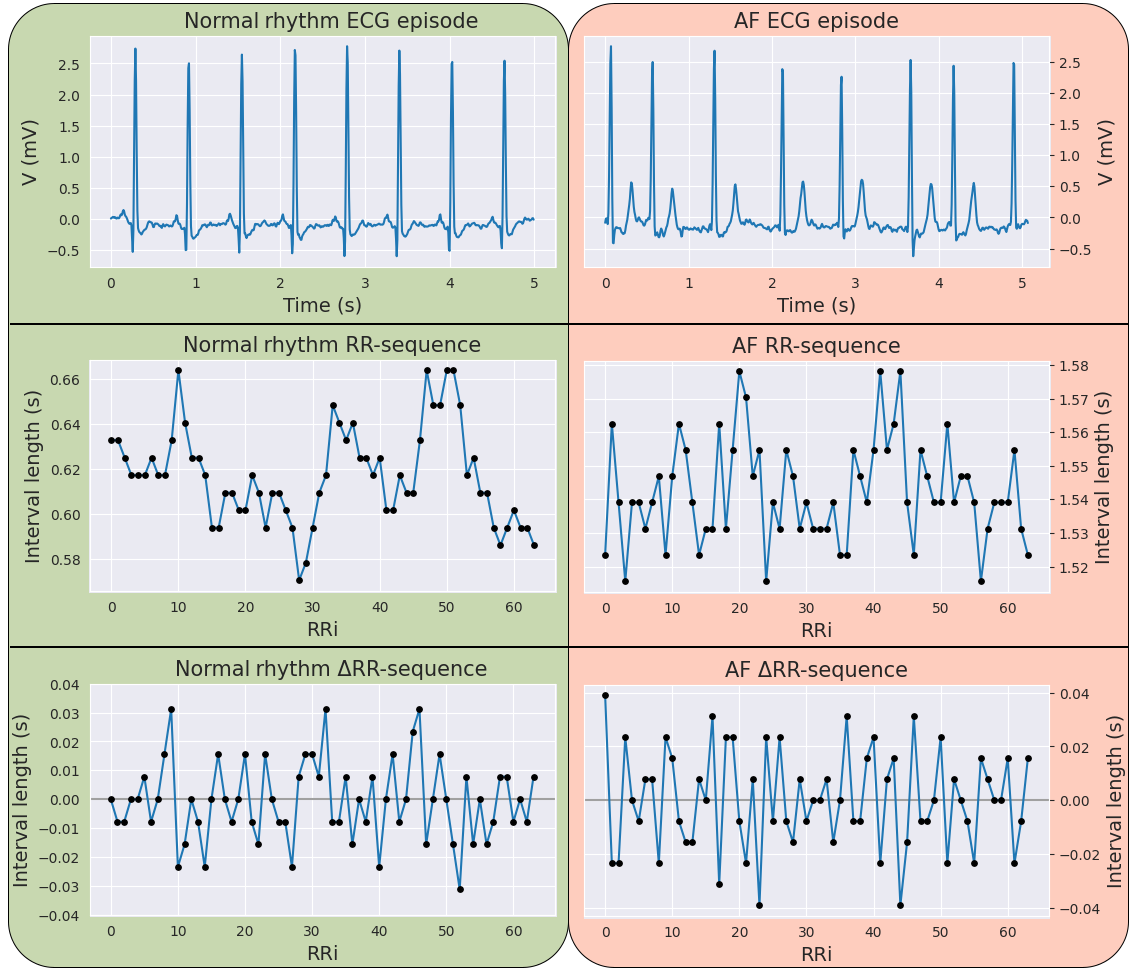}
    }
    \caption{Examples of normal rhythm (left) and AF episodes (right) represented as ECG, RR-interval, and $\Delta$RR-interval sequences.}
    \label{fig:rhythmplot}
\end{figure}

Automatic AF detection on ECG is one of the most studied problems in the field of biomedical engineering \cite{faust2020review}. To differentiate between normal rhythm and AF paroxysm, searching for irregularities in the time series of beat-to-beat RR-intervals or the absence of the P-wave in an ECG is recommended \cite{dash2009automatic}. Sometimes, RR-intervals classification (Fig. \ref{fig:rhythmplot}) is the preferred  AF detection approach due to the robustness to noise \cite{larburu2011comparative}, and the ability of inexpensive wearable devices to record heart rate \cite{georgiou2018can}. Methods and algorithms designed for the task usually involve pattern matching \cite{tateno2001automatic}, classical machine learning \cite{asgari2015automatic}, or deep neural networks \cite{xia2018detecting}.  

Naturally, neural networks achieve the best rhythm classification results. However, neural networks are computationally expensive and tend to overfit to specific datasets. This aspect complicates the implementation of neural network models in continuous health monitoring systems, such as mobile wearable devices or Holter monitors. Therefore, the need for lightweight AF detection methods continues to exist.

In 2023, Jiang Z. et al. introduced a novel method for text classification \cite{jiang2023low, jiang2022less}, which utilizes compressors such as gzip for capturing irregularity in data. Their method involved the calculation of compressor-based distances between data points and the use of a simple k-nearest neighbor ($k$NN) method to perform classification. The proposed method outperformed multi-million parameter transformer models and presented an exciting new development in the field of classifiers.

In this study, we applied Jiang's method to AF detection and binary classification between heart rhythms. Information irregularities captured by gzip compression are inherently data type agnostic, making the method intuitively applicable to the problem at hand. We compared the results between RR and $\Delta$RR measures and performed fivefold cross-validation of the method on the full MIT-BIH atrial fibrillation dataset. We investigated the effect of different $k$NN settings, searched for an optimal window length for classification, and considered different data types for sequence representation. In addition, we evaluated the classification performance under the few-shot learning setting.

\section{Methods and Materials}

\subsection{Dataset}
The research was conducted on the standard MIT-BIH Atrial Fibrillation Database (AFDB) from the open repository of biomedical signals PhysioNet \cite{goldberger2000physiobank}. AFDB consists of 25 Holter monitor ECG recordings of 10-h duration sampled at 250 Hz. The database is supplied with pre-calculated R-peak positions and verified rhythm annotations, making it one of the largest open datasets available for AF studies. 

The sequences of RR-intervals and $\Delta$RR-intervals were extracted and assigned class labels according to annotation markers. RR-interval is an interbeat interval, and $\Delta$RR-interval is the difference between successive RR-intervals. We explored both measures, as certain methods \cite{tateno2001automatic} have shown an improved accuracy with $\Delta$RR classification compared to RR. In addition, we investigated the classification performance for windows of different sequence lengths $M_{seq} = 32, 64$ and $128$ beats. The shorter sequence classifier was preferred as it allows the use of shorter ECG signals to detect AF paroxysms.

\subsection{Normalized Compression Distance and Data Types}
Jiang Z. et al. \cite{jiang2023low} suggested the use of normalized compression distance (NCD), defined as follows:
\begin{equation*}
NCD(x,y) = \frac{C(xy) - \textrm{min} \Bigl( C(x), C(y) \Bigr) }{\textrm{max}\Bigl( C(x), C(y) \Bigr) }.
\end{equation*}
Here, $x$ and $y$ are arbitrary sequences, $xy$ is their concatenation, and $C(\cdot)$ is the character length of a compressed sequence. This approach was proposed in \cite{li2004similarity, cilibrasi2005clustering}, with compressed length $C(x)$ serving as an approximation of Kolmogorov information distance \cite{kolmogorov1963tables}. 

The sequence $x$ in this paper is a series of RR-intervals or $\Delta$RR-intervals. AFDB signals are sampled at 250 Hz and interbeat intervals usually fluctuate between 300 and 3000 milliseconds, meaning an RR-interval here is a number of monitor counts typically in the range of 75 to 750. $\Delta$RR-interval is the difference between consecutive RR-intervals; therefore it's a value in the range of $-675$ to 675.

Both measures fit into 16-bit integer, therefore we chose it as the baseline data type. However, different data types present different bit distributions that may affect gzip compression. Hence, we also tested other data types: 8-bit, 16-bit and 32-bit, unsigned and signed integers. We investigated the effect of normalization of data to the entire integer range which in case of 16 and 32 bits leads to rather sparse bitwise representation. In case of 8 bits, the values were clipped to the range of $-750$ to 750 and quantized further to the integer range. For unsigned integers, the whole dataset was shifted forward by the absolute minimum value. We also tested 16-bit and 32-bit float divided by the sample rate, so that intervals were represented in seconds rather than monitor counts.

\subsection{k-Nearest Neighbor Classifier}
The matrix of NCD distances was calculated between the samples in the training and test sets. The next step was to use $k$NN to perform the classification of the test samples. Here, the $k$NN configuration was required to be considered. The choice of $k$ was important, as low values are too sensitive to noise while high values would be biased toward classes with higher number of observations in unbalanced datasets. We studied the algorithmic stability of the method to determine the impact of growing of $k$ on the classification performance.  
This choice of $k$ was explored for the method trained on the full dataset divided into different interval sequence lengths $M_{seq}$ as well as different data type representations of sequences. 

\subsection{Validation Approach on the Full Dataset}
We used fivefold cross-validation to verify the classification performance of the method on the full dataset. The AFDB dataset was split on a per-patient basis, so that the training and test populations for each of the five splits consisted of 20 and 5 recordings. This approach prevented data leaks and avoided the possibility of sequences extracted from one patient appearing in both the training and test sets, skewing the results.

The main metric we used to compare classifiers was macro F1-score \textendash \ an average of F1-scores per class. While this metric is not typically used in binary classification problems, we believe it is appropriate for the task due to varying degrees of class imbalance between cross-validation splits. This allowed us to choose the most balanced method configurations in terms of specificity and sensitivity.

\subsection{Few-shot Learning Approach}
To evaluate classification performance in few-shot learning setting, we took one of the train-test population splits created by fivefold routine and divided the training set into random even subfolds, each consisting of $n$ sequences of both classes. These subfold splits were created for each considered number of shots $n = 5,10,50,100,500,1000,2000$, and classification of the entire test set was performed with each subfold. With this approach, we could comprehensively assess the viability of few-shot learning for the problem.


\section{Results}
\subsection{Comparison between RR and $\Delta$RR}
NCD must show a clear distinction between normal rhythm and AF classes for the method to work. Therefore, between RR and $\Delta$RR measures, we need to select one with better defined interclass distances.

\begin{figure}[t]
    \centering
    \subfigure
    {
        \includegraphics[width=3.4in]{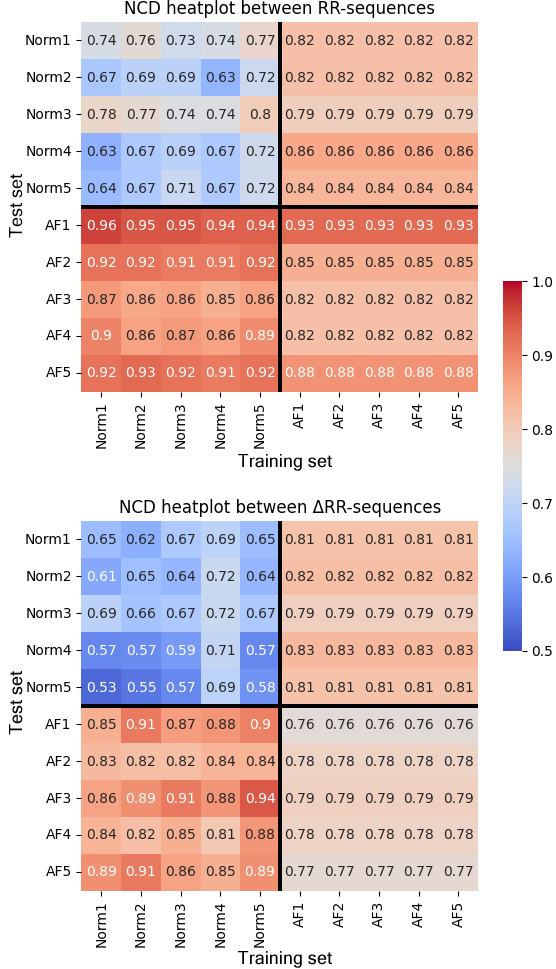}
    }
    \caption{Distance matrix of a small subset of the data. Training and test set were sampled from two different patients. The top graph shows the compression distances between RR-intervals, the bottom graph shows the compression distances between $\Delta$RR-intervals.}
    \label{fig:heatplot_small}
\end{figure}

We sampled a small training-test set of five normal rhythm sequences and five AF sequences from two different patients, and calculated their distance matrices for RR and $\Delta$RR measures. These matrices are shown in Fig. \ref{fig:heatplot_small}. As observed, the heatmap of $\Delta$RR matrix was more explicitly divided into four square clusters of interclass distances. 

\begin{figure}[t]
    \centering
    \subfigure
    {
        \includegraphics[width=3.4in]{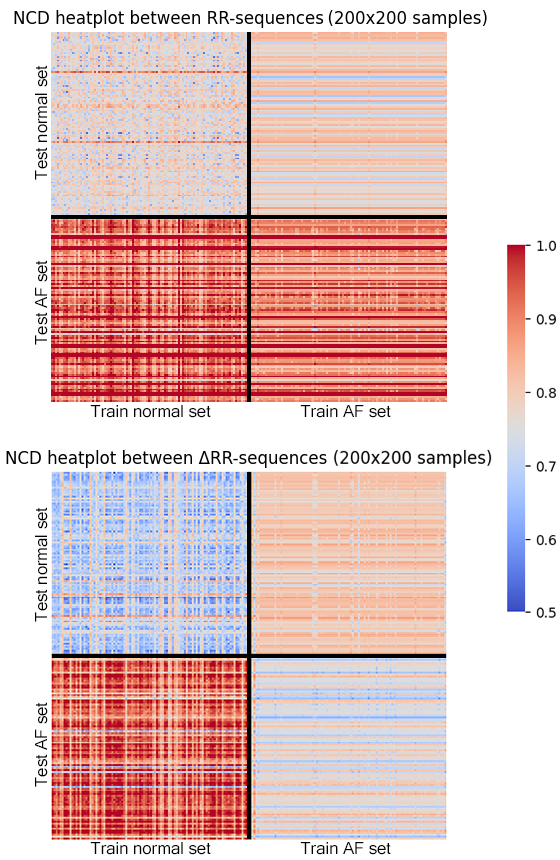}
    }
    \caption{Distance matrix of a large subset of the data. Training and test set were sampled from two non-intersecting groups of patients. The top graph shows the compression distances between RR intervals, the bottom graph shows the compression distances between $\Delta$RR intervals.}
    \label{fig:heatplot_big}
\end{figure}

The visual pattern persisted when we increased the sample size to 100 of both rhythm types and drew from a larger patient populations in both sets (Fig. \ref{fig:heatplot_big}). This result indicates that $\Delta$RR classification may yield better accuracy.

\begin{table}[h]
\caption{Fivefold classification results between RR and $\Delta$RR measures. ($M_{seq}$ = 64, $k$ = 501, $dtype$ = int16)}\label{tbl:deltarr_classification}
\centering
\begin{tabular}{|c|ccccc|c|}
\hline
Measure & Split 1 & Split 2 & Split 3 & Split 4 & Split 5 & Avg. \\ 
\hline
RR & 0.628 & 0.701 & 0.478 & 0.617 & 0.883 & 0.662 \\
\textbf{$\Delta$RR} & \textbf{0.951} & \textbf{0.920} & \textbf{0.816} & \textbf{0.899} & \textbf{0.941} & \textbf{0.905} \\
\hline
\end{tabular}
\end{table}

Classification results on the full dataset (Table \ref{tbl:deltarr_classification}; sequence length $M_{seq}$ = 64, $k$ = 501, int16) confirmed this finding. The macro average F1-scores of $\Delta$RR were higher and more stable for each cross-validation data split, as RR classification score dropped to 0.478 for split 3. Therefore, for the remainder of the paper, we present the results only for the $\Delta$RR classification as the clearly preferred approach.

\subsection{Classification on the Full Dataset}

\begin{table}[t]
\caption{5-fold classification results in macro average F1-scores. \\ ($M_{seq}$ is length of a sequence window, $k$ is neighbors in $k$NN, $dtype$ = int16)}\label{tbl:bigtable_classification}
\centering
\begin{tabular}{|l|ccccc|c|}
\hline
\multicolumn{7}{|c|}{$M_{seq}$ = 32}\\
\hline
$k$NN & Split 1 & Split 2 & Split 3 & Split 4 & Split 5 & Avg. \\ 
\hline
k=1	& 0.824	& 0.766 & 0.783 & 0.819 & 0.868 & 0.812 \\
k=5	& 0.889 & 0.813 & 0.830 & 0.863 & 0.918 & 0.863 \\
k=51 & 0.949 & 0.910 & 0.901 & 0.883 & \textbf{0.932} & 0.915 \\
k=101 & 0.953 & 0.931 & 0.910 & 0.884 & 0.927 & 0.921 \\
k=501 & 0.958 & 0.965 & \textbf{0.923} &	0.887 & 0.919 &	0.930 \\
\hline
\textbf{k=545} & \multirow{2}{*}{\textbf{0.960}} & \multirow{2}{*}{0.967} &	\multirow{2}{*}{0.922} & \multirow{2}{*}{\textbf{0.887}} & \multirow{2}{*}{0.917} &\multirow{2}{*}{\textbf{0.931}}  \\
 \emph{(best k)}& \multicolumn{5}{|l|}{} &  \\
 \hline
k=1001 & 0.959 & 0.972 & 0.913 & 0.887 & 0.913 & 0.929 \\
k=3001 & 0.958 & \textbf{0.982} & 0.900 & 0.887 & 0.907 & 0.927\\
\hline
\multicolumn{7}{|c|}{$M_{seq}$ = 64}\\
\hline
$k$NN & Split 1 & Split 2 & Split 3 & Split 4 & Split 5 & Avg. \\ 
\hline
k=1 & 0.793 & 0.611 & 0.722 & 0.769 & 0.826 & 0.744 \\
k=5	& 0.853 & 0.591 & 0.728 & 0.803 & 0.885 & 0.772 \\
k=51 & 0.938 & 0.676 & 0.778 & 0.869 & 0.941 & 0.840 \\
k=101 & 0.945 &	0.744 & 0.807 & 0.881 &	\textbf{0.950} &	0.865 \\
k=501 & \textbf{0.951} & 0.920 & 0.816 & 0.899 & 0.941 & 0.905 \\
k=1001 & 0.950 & 0.971 & 0.825 & 0.903 & 0.932 & 0.916 \\
k=3001 & 0.943 & 0.987 & 0.907 & \textbf{0.908} & 0.914 & 0.932 \\
\hline
\textbf{k=4779} & \multirow{2}{*}{0.936} & \multirow{2}{*}{\textbf{0.988}}  & \multirow{2}{*}{\textbf{0.955}} & \multirow{2}{*}{0.901} & \multirow{2}{*}{0.902}  & \multirow{2}{*}{\textbf{0.937}}  \\
 \emph{(best k)}& \multicolumn{5}{|l|}{} &  \\
\hline
\multicolumn{7}{|c|}{$M_{seq}$ = 128}\\
\hline
$k$NN & Split 1 & Split 2 & Split 3 & Split 4 & Split 5 & Avg. \\ 
\hline
k=1 & 0.613 & 0.375 & 0.553 & 0.635 & 0.620 & 0.559 \\
k=5	& 0.677 & 0.322 & 0.528 & 0.677 & 0.662 & 0.573 \\
k=51 & 0.727 & 0.370 & 0.434 & 0.746 & 0.744 & 0.604 \\
k=101 & 0.744 &	0.414 &	0.415 & 0.758 &	0.769 &	0.620 \\
k=501 & \textbf{0.889} & 0.818 & 0.817 & 0.843 & \textbf{0.861} & 0.846 \\
k=1001 & 0.867 & 0.985 & 0.939 & 0.871 & 0.829 & 0.898 \\
\hline
\textbf{k=1343} & \multirow{2}{*}{0.853} & \multirow{2}{*}{\textbf{0.990}}  & \multirow{2}{*}{\textbf{0.972}} & \multirow{2}{*}{\textbf{0.873}}  & \multirow{2}{*}{0.813} & \multirow{2}{*}{\textbf{0.900}}  \\
 \emph{(best k)}& \multicolumn{5}{|l|}{} &  \\
 \hline
k=3001 & 0.808 & 0.988 & 0.969 & 0.864 & 0.771 & 0.880 \\
\hline
\end{tabular}
\end{table}

The full dataset was divided into $\Delta$RR sequences of length $M_{seq} = 32, 64, 128$, and we examined the classification performance of gzip method for each mode. Table \ref{tbl:bigtable_classification} presents detailed results for a selection of $k$-values. According to our investigation, $M_{seq} = 32$ and $M_{seq} = 64$ were the preferred lengths and reached the total average F1-score of 0.931 and 0.937 respectively, while for $M_{seq}=128$ the best result was 0.900. This result aligned with the original work \cite{jiang2023low}, as gzip has shown better classification of datasets with shorter texts. Classification could be slightly more accurate for $M_{seq} = 64$, but in the case of $M_{seq} = 32$, the classifier as more stable for lower number of neighbors $k$.  

\begin{table}[h]
\caption{Best classifiers in terms of specificity and sensitivity among 5-fold splits. Best sensitivity and specificity are shown for the best split.}\label{tbl:best_classifiers}
\centering
\begin{tabular}{|cc|cc|cc|}
\hline
$M_{seq}$ & $k$NN & Avg. Sens. & Avg. Spec. & Best Sens. & Best Spec. \\
\hline
32 & \textbf{k=545} & 93.6\% & \textbf{93.5\%} & 97.2\% &  95.3\%  \\
64 & k=4779 & 97.1\% & 91.7\%  & \textbf{99.8\%} & 97.6\%  \\
128 & k=1343 & \textbf{98.6\%} & 86.1\%  & 99.8\% & \textbf{97.7\%}   \\
\hline
\end{tabular}
\end{table}

Table \ref{tbl:best_classifiers} compiles best classifiers and their average sensitivity and specificity between cross-validation splits. Classification of longer sequences yields higher average sensitivity, but lower average specificity. Therefore, the choice of window $M_{seq}$ can be made according to practical needs: classification on shorter sequences requires shorter ECG recordings, but longer sequences may result in fewer false negatives.


\subsection{Data Type Examination}
In addition to the baseline \textbf{int16} data type, we examined the classification performance for other integer data types (int8, uint8, uint16, int32, uint32) and real data types (float16, float32). Integer data types were tested as original values or normalized to integer range. In case of unsigned integers, values were shifted toward right by the absolute dataset minimum. In case of 8-bit int, $\Delta$RR-intervals did not fit in the variable range; hence, the values were clipped and quantized to integer range. Float values were presented as seconds rather than monitor counts. The window $M_{seq}$ was fixed to 64 for these experiments.

\begin{table}[h]
\caption{Classification performance between data types in terms of total average F1-scores between five cross-validation splits ($M_{seq}$ = 64)}\label{tbl:datatypes}
\centering
\begin{tabular}{|c|cccccc|c|}
\hline
\multicolumn{8}{|c|}{\textbf{Original values (monitor counts)}}\\
\hline
dtype & k=1 & k=5 & k=51 & k=101 & k=501 & k=1001 & \textbf{best k} \\
\hline
\textbf{int16} & 0.744 & 0.772 & 0.840 & 0.865 & 0.906 & 0.916 & \textbf{0.937} \\
& & & & & & & \textit{k=4779} \\

uint16 & 0.806 & 0.842 & 0.863 & 0.870 & 0.892 & 0.904 & 0.925 \\
 & & & & & & & \textit{k=3781} \\

\hline

int32 & 0.674 & 0.664 & 0.697 & 0.726 & 0.811 & 0.855 & 0.935 \\
& & & & & & & \textit{k=5831} \\
uint32 & 0.777 & 0.820 & 0.873 & 0.884 & 0.902 & 0.903 & 0.934 \\
& & & & & & & \textit{k=5091} \\
\hline
\multicolumn{8}{|c|}{\textbf{Normalized to integer range (monitor counts)}}\\
\hline
dtype & k=1 & k=5 & k=51 & k=101 & k=501 & k=1001 & \textbf{best k} \\
\hline
int8 & 0.847 & 0.866 & 0.854 & 0.867 & 0.897 & 0.903 & 0.924 \\
 & & & & & & & \textit{k=5133} \\
uint8 & 0.843 & 0.872 & 0.891 & 0.904 & 0.921 & \textbf{0.922} & 0.924 \\
 & & & & & & & \textit{k=1363} \\
\hline
int16 & 0.808 & 0.837 & 0.852 & 0.852 & 0.845 & 0.867 & 0.931 \\
 & & & & & & & \textit{k=6891} \\
uint16 & 0.729 & 0.737 & 0.758 & 0.768 & 0.799 & 0.828 & 0.926 \\
 & & & & & & & \textit{k=6985} \\
\hline
int32 & 0.794 & 0.803 & 0.788 & 0.777 & 0.763 & 0.759 & 0.805 \\
& & & & & & & \textit{k=7} \\
uint32 & 0.792 & 0.815 & 0.812 & 0.807 & 0.795 & 0.794 & 0.815 \\
& & & & & & & \textit{k=7} \\
\hline
\multicolumn{8}{|c|}{\textbf{Real values (s)}}\\
\hline
dtype & k=1 & k=5 & k=51 & k=101 & k=501 & k=1001 & \textbf{best k} \\
\hline
float16 & 0.781 & 0.818 & 0.870 & 0.886 & 0.897 & 0.894 & 0.921 \\
& & & & & & & \textit{k=4551} \\
float32 & \textbf{0.874} & \textbf{0.904} & \textbf{0.914} & \textbf{0.915} & \textbf{0.906} & 0.903 & 0.919 \\
& & & & & & & \textit{k=3035} \\
\hline
\end{tabular}
\end{table}

Table \ref{tbl:datatypes} presents the total average F1-scores between five cross-validation splits for listed data types along different $k$-values. Baseline int16 reached the best overall accuracy. However, many of the tested data types showed better stability of the method for lower values of $k$.  Of particular interest was float32 that showed the best stability and classification accuracy for $k \in [1\textup{--}501]$. This result might be related to dense utilization of bits in binary floating point representation of $\Delta$RR-intervals, which tend to oscillate around 0. The most stable integer data type was uint8, which required shifting, clipping, and quantization of the dataset, resulting in dense bit utilization as well. The worst results were found for int32 and uint32 normalized to the integer range, suggesting that very sparse bit allocation leads to worse irregularity detection with gzip.

\subsection{Few-shot Learning Evaluation}
We took the first train-test patient split previously created by fivefold cross-validation routine on the full dataset, and we further divided the training set into random subfolds that included $n$ examples of both rhythm classes. For the window $M_{seq}=32$ and the number of shots $n = 5,10,50,100,500,1000,2000$ this amounted to $2649,1324,264,132,26,13,6$ training subfolds, respectively. We performed classification of the whole test set (which included 6736 sequences) using each subfold with the number of neighbors $k$ fixed to $2*\sqrt{n}$ rounded to the closest odd number.

\begin{figure}[h]
    \centering
    \subfigure
    {
        \includegraphics[width=3.4in]{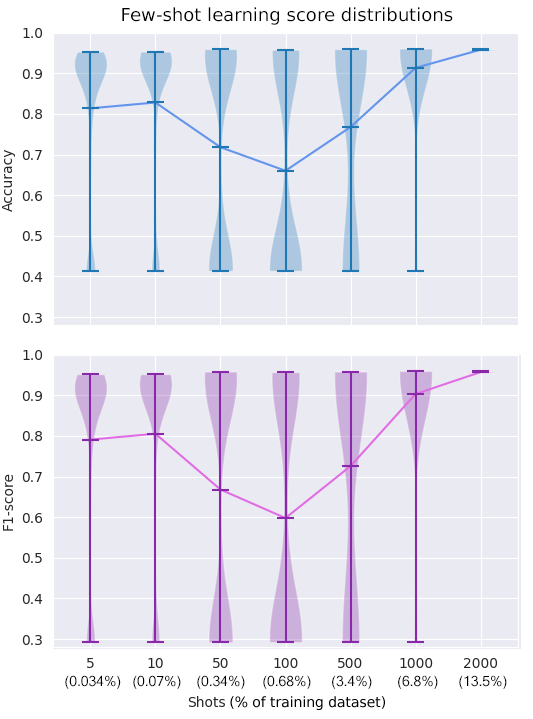}
    }
    \caption{The plots showing distributions and the change in scores with the increasing number of shots under few-shot learning setting. The violinplots represent distributions and the lineplot between them shows changing average score. The top graph shows accuracies and the bottom graph shows F1-scores.}
    \label{fig:violinplots}
\end{figure}

Results in terms of accuracy and F1-scores are presented as violinplots in Fig. \ref{fig:violinplots}. As observed, the violinplots were very top and bottom heavy, meaning that each subfold either succeeded and performed classification with excellent accuracy, or assigned the same class to the entire test set. The ratio of "faulty" subfolds was higher for $n = 50,100,500$ than for $n = 5,10$, which is reflected in decreasing average scores with increasing number of shots. For $n = 1000$ and $2000$ the average F1-scores jumped back to 0.903 and 0.957. From this result, we could conclude that the entire test set can be reliably classified within the range of 6\% to 13\% of the training  set. The existence of "faulty" subfolds suggests that the exclusion of noisy data points can increase the stability of few-shot learning and reduce the required training sample down to a fraction of a percent. Overall, the classification method showed robust generalization capabilities.

\section{Discussion and related works}
In this paper, we investigated the efficiency of compressor-based classification for the detection of atrial fibrillation on ECG using sequences of interbeat intervals. We identified $\Delta$RR classification as the preferred approach compared to RR classification. For the window size 32, the method achieved an average sensitivity of 93.6\% and an average specificity of 93.5\% on the full MIT-BIH Atrial Fibrillation Database. For the wider window size of 64, the method showed an average sensitivity of 97.1\% and an average specificity of 91.7\%. Our cross-validation routine divided the patients into separate training and test sets, preventing data leaks.

Classification based on gzip compression has following advantages over other popular AF detection algorithms: 
\begin{itemize} 
\item Tateno's algorithm \cite{tateno2001automatic} is one of the classical works in the field, as they proposed the use of $\Delta$RR sequences for classification by histogram pattern matching. They achieved a sensitivity of 94.4\% and specificity of 97.2\% for window size $M_{seq}$ = 100. However, their classification performance dropped significantly for shorter sequences and they did not provide cross-validation results on AFDB.

\item Asgari et al. algorithm \cite{asgari2015automatic} employs stationary wavelet transform and support vector machine to detect AF using surface ECG signal. Their method uses only a 10-s ECG segment for classification, does not require R-peak detection and achieves a sensitivity and specificity of 97.0\% and 97.1\%, respectively. However, they use a different validation routine (twofold instead of fivefold), and our results on the best splits (99.8\% sensitivity, 97.6\% specificity for $M_{seq} = 64$) outperformed theirs.

\item Neural network classifiers such as \cite{xia2018detecting} achieve some of the best  results in terms of accuracy (98.7\% sensitivity, 98.9\% specificity). However, neural network methods are quite computationally expensive and prone to overfitting.

\end{itemize}

We tested different data type representations of sequences and identified that $\Delta$RR intervals can be represented as either seconds (float32) or quantized down to a lower sampling rate (uint8) with improved stability relatively to the number of neighbors $k$. Classification results under the few-shot learning setting suggest that only a fraction of the training set (6\%-13\%) is required to perform the classification of the test set. Taken together, these results assure that the method does not require an ECG recorded at a specific sampling rate or a large amount of data to perform classification. Since gzip compression is often optimized on hardware level \cite{abdelfattah2014gzip}, we believe that gzip-based classification may be more computationally efficient than methods based on neural networks. Compressor-based classification is especially promising for devices with low energy consumption: portable ECG monitors, wearable electronics, telehealth and remote patient monitoring.   

\section{Conclusion}
In this paper, we studied classification with normalized compression distance for the task of the AF detection in electrocardiographic data. Using $\Delta$RR-interval series, we achieved the classification performance (average sensitivity of 97.1\% and an average specificity of 91.7\%, best sensitivity of 99.8\%, best specificity of 97.6\%) approaching specialized algorithms for AF detection. 

According to our results, the baseline int16 data type achieves the best possible classification performance. However, float32 and uint8 are more stable and less sensitive to the number of neighbors $k$, which may render them more suitable for practical tasks. Classification on shorter sequences with the window length $M_{seq} = 32$ yielded slightly better specificity and stability relative to $k$. Classification with the window length $M_{seq} = 64$ brought higher sensitivity and overall F1-score. The best of choice of $k$ varied greatly between different data types, and for the window length $M_{seq} = 64$ it generally tended to be 4\% to 20\% of the training set size. Classification with gzip compression does not work with RR-intervals and instead requires $\Delta$RR-intervals.

Originally, compressor-based classification was proposed for a series of discrete states, such as texts, discrete random sequences, \textit{etc.} Our study shows, that normalized compression distance is also suitable for classification of quantized continuous stochastic sequences. We expect good performance of normalized compression distance in other ECG-processing tasks and biomedical applications.

\bibliographystyle{./IEEEtran}
\bibliography{bibliography.bib}

\section*{Author information}

\begin{itemize}
    \item Nikita Markov$^{1,2,3}$, ns.markov@urfu.ru
    \item Konstantin Ushenin$^{1,2,3}$, konstantin.ushenin@urfu.ru
    \item Yakov Bozhko$^{1}$, yakov-bozhko@yandex.ru
    \item Olga Solovyova$^{2,3}$, o-solovey@mail.ru
\end{itemize}

\begin{enumerate}
    \item Ural State Medical University, Yekaterinburg, Russia
    \item Ural Federal University, Yekaterinburg, Russia
    \item Institute of Immunology and Physiology UrB RAS, Yekaterinburg, Russia
\end{enumerate}

Acknowledgments: State task to USMU No. 730000F.99.1.BV10AA00006 ``Development of an integrated approach to personalized diagnosis, therapy and prevention of supraventricular cardiac arrhythmias of autonomic genesis''

\end{document}